# Ultrasonic Defect Modification in Irradiated Silicon.

## Lucien Cremaldi, and Igor Ostrovskii

Department of Physics and Astronomy, University of Mississippi, University MS 38677.

## Abstract

It is shown for the first time, that room temperature Ultrasonic Defect Manipulation (UDM) can significantly reduce the concentration of radiation defects in high resistivity silicon. Secondary Ion Mass Spectroscopy revealed that oxygen- and hydrogen- related chemical reactions in silicon are likely to occur under UDM at room temperature. Ultrasonically stimulated chemical reactions in solids can be an important source of energy, which is required for UDM.

## 1. Introduction.

The understanding of radiation defects in solids is an important part of contemporary physics. It is well known, that nuclear radiation can generate point defects in silicon [1], cause its amorphization, in case of ion-irradiated silicon [2], change silicon optical properties by high-energy-electron irradiation [3], and affect electrical characteristics of Si-based detectors [4]. Silicon is used in nuclear radiation detector and calorimeter systems and is exposed to nuclear radiation of different origin. As a result, an increase in concentration of point defects and their associates leads to degradation of silicon properties and performance. For example, in silicon diverse point defects like donors, acceptors, recombination centers, and diverse complexes of imperfections, including those containing oxygen, are altered in their concentration.

There are only a few ways of enhancing the radiation hardness of solids. Traditional annealing by applying a relatively high temperature of about 900 $^0$C to the sample is limited in application to initially prepared thin chips or bulk samples. Thermal treatment affects not only oxygen and carbon concentrations in Si but also all defect subsystems.

To improve material performance, a defect manipulation  (sometimes referred to as "defect engineering") approach was developed during last years [5-8]. This relatively new method may use an excessive concentration of dopants, mainly oxygen in case of silicon [5], or low temperature treatment of crystalline solids by acoustical vibrations of ultrasonic frequencies [6 - 8].

In this work we describe the experimental results on ultrasonic defect manipulation in silicon. The goal of this research is to show the possibility of radiation defects manipulation in single crystal of silicon by ultrasound at room temperature. Based on our experiments, we discuss possible physical mechanisms of acoustic wave influence on point defects, and in doing so address the possibility of curing radiation damage in silicon crystals by ultrasound.   We call this approach "ultrasound defect manipulation" (UDM). We may note, UDM can be applied not only to a single crystals damaged by high-energy radiation but also to real detectors or detector array systems.

## 2. Ultrasonic Defect Manipulation.

The Ultrasonic Defect Manipulation (UDM) approach consists of ultrasonically altering the concentration of point defects, their associates, and their space location within the crystal bulk, including the near surface region. This last in particular is very important for metal-semiconductor contact parameters, which in turn influence the nuclear radiation detector performance. In our experiments, an acoustic wave of megahertz frequency range is applied to the irradiated samples. The experiments are performed at room temperature, but during ultrasonic application, the sample temperature is slightly elevated. Under the UDM treatment, the maximum temperature increase is about 30 degrees above initial room temperature. This extra thermal energy means a partial conversion of ultrasonic energy into heat, which is useful effect for an activation of point defect migration. We note the temperatures of near 300K and higher up to about 550K can be defined as low temperatures for temperature-based annealing processes.

The samples have the form of a parallelepiped and are cut off from a massive piece.  The dimensions are about 5 x 5 mm$^2$ for the end faces, and the length is about (10–12) mm. In our experiments, an ultrasonic wave of MHz-frequency range is applied to the samples with the help of attached ultrasonic transducers, which are made of PZT-ceramic. To generate acoustic wave, an rf-voltage (V) is applied to the transducer.

Two experimental techniques are employed in order to detect point defect migration under the room temperature ultrasonic treatment: 1) electrical conductivity measurements, and 2) Secondary Ion Mass Spectrometry (SIMS). In the case of silicon, electrical conductivity is an important parameter to measure. It is strongly altered by nuclear radiation. The samples of Czochralski (Cz) grown p-type silicon were first irradiated by neutrons. To measure electric conductivity, two metal ohmic contacts are deposited on two opposite faces of the sample. We then measure a change in electric conductivity of Si after UDM-treatment.

A kind of direct measurements of atomic migration due to UDM was performed with the help of a secondary ion mass spectroscopy technique. This series of experiments are very important to answering a general question: "Is ultrasound able to stimulate atomic migration in silicon for more heavy species than hydrogen?". Hydrogen itself can effectively migrate under ultrasound through the silicon lattice even in the case of polycrystalline phase [9]. Some publications report so called redistribution of point defects under ultrasonic action in different materials like CdS [10] and polycrystalline silicon thin films [9], InSb, etc. [6 - 8], but the case of silicon remains unclear. In Si-single crystal, potential barriers for interstitial atoms migration are respectively high, and that is why a direct observation of impurities migration stimulated by ultrasound at room temperature is a distinct physical problem. To answer this question, the SIMS-technique is an important experimental method.

### 3. Electrical conductivity measurements.

The experiments are carried out with the samples irradiated by neutron radiation. Before UDM, the samples of p-type Cz-Si had an initial electrical resistivity of $10^5$ Ohmcm. The conductivity of Si is very sensitive to concentration of the crystal defects; hence a variation of sample resistance can reveal any change in the defect subsystem including concentration or electrical activity. We apply to the Si samples an ultrasound of 2.16 MHz frequency during 3 to 5 minutes. The rf-voltage amplitude V is in the range from 0 to 27 volts effective. The electrical conductivity is measured after each UDM application; the successive UDM applications of 3 minutes duration each are performed with gradually increasing rf-voltage amplitude V. Hence, each point in Fig. 1 corresponds to such measurement. As it is clear from Fig. 1, as higher acoustic amplitude is applied a higher change in conductivity is observed. The experimental observation under excitation level higher than 22 volts shows significant increase in conductivity, which can be associated with atomic/ionic migration stimulated by ultrasound and subsequent reduction in concentration of the radiation defects, which were generated in our samples by neutron radiation prior to UDM.

### 4. SIMS measurements.

The samples of bulk single Si crystal for SIMS-investigation are made from n-type material grown by Czochralski method. The samples were cut from a central part of the initial crystal with final dimensions of about 3.5 x 4.5 x 11 mm$^3$. The specific resistance of our samples was 16 Ohm·cm. In such n-Cz-Si, one can expect a relatively large concentration of oxygen and carbon.

Three SIMS spectra were taken initially from three different samples of this series. These initial spectra reveal similar features. After that, two other samples (Cz-n-Si-2K and Cz-n-Si-3K) are treated by ultrasound at room temperature undergoing our UDM procedure. One sample (Cz-n-Si-1K ) is not ultrasonically treated, and it is used for SIMS calibration during our different series of measurements. After UDM, the SIMS spectra are again measured for all the three samples. No changes are detected from the control sample Cz-n-Si-1K, but the significant changes are revealed after UDM from the two treated samples (Cz-n-Si-2K and 3K). The differential SIMS-spectra are presented in Fig. 2. It shows the difference between the numbers of mass counts after UDM and the number of same mass counts from the initial sample that is before UDM treatment. The change in SIMS counts from the UDM treated sample is given in percentages with respect to initial its spectrum.

For statistical reasons, we took a number of spectra. From each sample at least 5 spectra are measured before UDM and then also 5 spectra after UDM. There are some insignificant variations in the different experimental series, which we explain by inevitable influence of ionic etching of the samples during the measurements.

Nevertheless all the main features in the spectra remain the same for the different series of measurements. Fig. 2 shows the difference calculated from two successive measurements taken right before (initial) and after UDM treatment of the Cz-p-Si-2K sample.

## 5. Discussion.

The main increase in electrical conductivity_is observed under UDM at V > 12 volts, despite some small electrical conductivity change observed at lower amplitudes (V < 8 Volts). It is also observed the residual change in electrical conductivity taken 90 minutes after UDM (Fig. 1, plot 3), which means a positive UDM or radiation defects annealing by ultrasound.   In a logarithm scale, after a small-applied voltage of about 2 Volts, the observed dependence of conductivity change versus ultrasound amplitude becomes close to a linear reliance. The linear dependence suggests an exponential function describing the UDM influence on electrically active defects in Si. According to thermodynamic consideration, a concentration of electrically charged vacancies and interstitial atoms is a function of external pressure **P**.

$$n_+(\mathbf{P}) = n_-(\mathbf{P}) = (N_+ N_-)^{1/2} \exp[\frac{S_{F+} + S_{F-}}{2k} - \frac{E_+ + E_-}{2kT} - \frac{P(V_{F+} + V_{F-})}{2kT}] \qquad (1)$$

Where: $n_{+,-}$ is concentration of point defects having positive or negative electric charge; $N_{+,-}$ is concentration of corresponding regular sites in a crystal lattice; $S_{F+,-}$ and $E_{+,-}$ are the entropy and energy of defect formation. In case of UDM, the pressure exerted by ultrasound on a microscopic volume of a sample is a Sine-function of time

**P = P₀Sin(ωt)**, but due to the exponential character of  the amplitude dependence given by equation (1), the net result of UDM is not zero. Since a simple radiation defect, let us say an interstitial atom, could get back to its native crystallographic site in one half cycle of the acoustic wave, it also will be at its regular site in the next, opposite half cycle of the wave. The last approach is possible because ultrasound itself possesses much less energy than the nuclear radiation, which originally has produced this defect. From a theoretical point of view, the situation mentioned above can be analyzed by substituting the acoustically exerted pressure **P = P₀Sin(ωt)** by its module value or its magnitude

**IP₀Sin(ωt)I**. Thus the UDM treatment can lead to a decrease in defect concentration.

<u>Physically</u> it means that in our UDM-experiments with neutron-irradiated p-type Cz-Si, an ultrasonically stimulated diffusion — somewhat resembles a thermally-activated type of point defect diffusion.

A quantum mechanical picture of self-diffusion in silicon is recently described in [11]. The measured energies of migration of Si-vacancies in different charge states are known from 0.18 eV (double-positively charged) to 0.45 eV (neutral) [12 - 14]. The migration energy is much smaller for some impurities in silicon.  For different complexes of hydrogen, it is measured to be as low as 6.5 meV and 37.4 meV  [15]. At such small activation energies, ultrasound at room temperature is a stimulating diffusion factor even under its respectively low amplitudes.

Based on our experimental data and published activation energies, we can conclude the physical micro-mechanism of UDM in neutron-irradiated silicon might be connected to the primary diffusion of impurity atoms and ions and not of silicon vacancies. We have to note that different physical mechanisms of UDM action on solids result in local pressure and temperature increases, local dynamic deformations, and local electric fields due to a deformation potential. Nonlinear mechanisms increase the density of phonons, mainly near Brillouin zone boundary.   All mentioned above processes together could stimulate atomic migration in solids at room temperature.

<u>Analysis of SIMS spectra</u> is based on Fig. 2. The first simple conclusion comes from the changes in the peak amplitudes after UDM. Lighter elements [23]Na and [39]K are found to significantly increase in abundance after UDM treatment. The differential spectrum of Fig. 2  also reveals a new important peculiarity. The hydrogen and oxygen count drop by 25% and about 50%, respectively. Also the count for mass M = 17 a.m.u. is dropping, which can be associated with hydrogen-oxide (OH) group.  At the same time, the counts for the masses M = 45 a.m.u. (most probable is SiOH complex), and M = 44 a.m.u. (SiO group) increase by about 75% and 38%, respectively. These changes can be only explained by <u>chemical reactions inside Si</u> including its near-surface region during the UDM application. We do not see any alternative explanation. Consequently, such a kind of ultrasonically stimulated chemical reaction in solids is an important source of locally delivered energy, which in turn can stimulate a diffusion of doping/impurity atomic species. Among them, we admit an increase in iron counts and decrease in titanium counts.

Thus one can conclude ultrasonically stimulated chemical reactions in solids can be a source for locally deliberated energy of the order of electron volts, which is enough to promote a jump of interstitial ion/atom or vacancies to another close location.

An attempt to locate similar findings in the literature including chemical reactions stimulated by ultrasound in solids did not give a positive result. Among other close research efforts, we can mention studies describing a shock-wave interaction with microscopic voids in solids, which can lead to "hot spot" initiation chemistry [16], ultrasound stimulated dissociation of Fe-B pairs in silicon [17], and acoustically stimulated phase transitions in $PbI_2$ [18]. An increase in counts for sodium and potassium after ultrasonic treatment of Si was reported in [19], where SIMS spectra were detected for a narrow range of masses from M = 23 to 44 a.m.u.

To the best of our knowledge, the results given in this work for the whole spectrum of masses from M = 1 a.m.u. (Hydrogen) to M = 56 a.m.u. (Iron), differential SIMS spectrum, and subsequent conclusions on ultrasonically stimulated chemical reactions in silicon are presented for the first time.

For the reason of generality, we should mention another possible mechanism of locally occurring energy inside a solid under UDM. It has been shown experimentally, that ultrasonic treatment of semi-conducting materials can lead to a transfer of electrical charge between point defects and changes in the electronic subsystem [6, 8]. In case of dislocation free silicon, this results in an alteration of semiconductor photoelectric properties [20]. It also has been shown [21], that electronic excitation can in turn stimulate an atomic migration in a semiconductor. So we cannot exclude processes of this type in our experiments.

On the other hand, a detail analysis of the physical mechanisms responsible for atomic migration in our experiments is out of the scope of the present work, which primarily intends to show experimentally an existence of ultrasonically stimulated atomic migration for the particular important case of irradiated silicon.

## 6. Conclusions.

1. It is shown experimentally for the first time, that room temperature Ultrasonic Defect Manipulation of silicon containing radiation defects can significantly reduce the concentration of radiation defects in high resistive material. This conclusion is consistent with electrical conductivity measurements.

2. Secondary Ion Mass Spectroscopy measurements reveal significant ultrasonically stimulated atomic migration of relatively light elements like carbon, sodium, potassium, titanium and iron.

3. The analysis of ultrasonically caused changes in the SIMS spectra for hydrogen, oxygen and masses containing OH-groups permits us to conclude that oxygen- and hydrogen-related chemical reactions in silicon are likely to occur under UDM at room temperature.

4. The physical mechanisms responsible for ultrasonically stimulated atomic migration can be connected to atomic/ionic diffusion, activated by acoustic wave, and by important secondary processes of the type of chemical reactions and electrical charge transfer in solids while acoustic stress is present.

## 7. References:

[1] B.G. Svenson, C. Jagadish, And J.S. Williams, Phys. Rev. Let. **71**, 1860 (1993).

[2] D.N. Seidman, R.S. Averback, P.R. Okamoto, and A.C. Baily, Phys. Rev. Lett. **58,** 900 (1987)

[3] C.S. Chen, J.C. Corelli, and G.D. Watkins, Phys. Rev. B, **5**, 510 (1972).

[4] M. Moll, E. Fretwurst, G. Lindström. Nuclear Instruments and Methods in Physics Research. Sec. **A**, p. 87-93 (1999).

[5] G. Lindström et al. (RD48), "Radiation Hard Silicon Detectors – Developments by the RD48 (ROSE) Collaboration", 4-th STD Hiroshima Conf., March 2000. RD48 3-rd Status Report, CERN/LHCC 2000-009, Dec. (1999).

[6] I.V.Ostrovskii, Ju.M.Khalack, A.B.Nadtochii, H.G. Walther. Solid State Phenomena. **67-68**, 497-502 (1999).

[7] I. Dirnstorfer, W. Burkhardt, B. K. Meyer, S. Ostapenko, and F. Karg. Solid State Communications, **116**, 87-91 (2000). [Solid State Phenomena, **85-86,** 317 (2002).]

[8] B. Romanjuk, D. Krüger, V. Melnik, V. Popov, Ya. Olikh, V. Soroka, O. Oberemok. Semiconductors Physics, Quantum Electronics & Optoelectronics, **3**, N1, 15 (2000).

[9] S.Ostapenko, L.Jastrzebski, J.Lagowski amd R.Smeltzer. Appl.Phys.Lett., **68**, 2873 (1996).

[10] Ostrovskii I.V., Rozko A. Acoustic Redistribution of Defects in Crystals. Sov. Phys. Solid State. **26**, N12, 2241 (1984).

[11] Anna Jääskeläinen, Luciano Colombo, and Risto Niemen, Phys. Rev. B, **64**, 233203 (2001).

[12] G.D. Watkins, J. Phys. Soc. Jpn. Suppl. II. **18**, 22 (1963).

[13] J.R. Troxell and A.P. Chatterjee. Inst. Phys. Conf. Ser. **46**, 16 (1979).

[14] J.R. Troxell and G.D. Watkins, Phys. Rev. **B 22**, 921 (1980).

[15] M. Suezava, Y. Takada, T. Tamano, R. Taniguchi, F. Hori, and R. Oshima, Phys. Rev. **B 66**, 155201 (2002).

[16] B.L. Holian, T.C. Germann, J-B. Mailet, and C. T. White, Phys. Rev. Lett., **89**, N28, 285501 (2002).

[17] S. S. Ostapenko and R. E. Bell, J. Appl. Phys. **77**, 5458 (1995).

[18] M.M. Bilyi, I.S. Gorban, I.M. Dmitruk, I.M. Salivonov, and I.V. Ostrovskii, Low. Temp. Phys, **24**, N8, 609 (1998).

[19] I.V. Ostrovskii, A.B. Nadtochij, A.A. Podolyan, Semiconductors. **36**, N4, 367 (2002).

[20] I.V. Ostrovskii, A. B. Nadtochii, L.P. Steblenko, A.A. Podolian, 2001 IEEE Ultrasonics Symposium, Oct. 07-10, Atlanta, USA, Proceedings, **1**, 401 (2001).

[21] V.S. Vavilov, Uspeh. Phys. Nauk (Russian issue). **167**, N4, 407 (1997)

**FIGURE CAPTIONS**

Fig. 1. Influence of UDM on electrical conductivity taken from a neutron irradiated Czchohralski p-type silicon. Plot 1 – measurements are made 30 min after UDM, 2 – 60 min   after UDM, 3 – 90 min after UDM.

Fig. 2. Effect of UDM on SIMS spectrum from Cz-Si. Negative values – counts became lower after UDM, positive values – counts became higher after UDM.

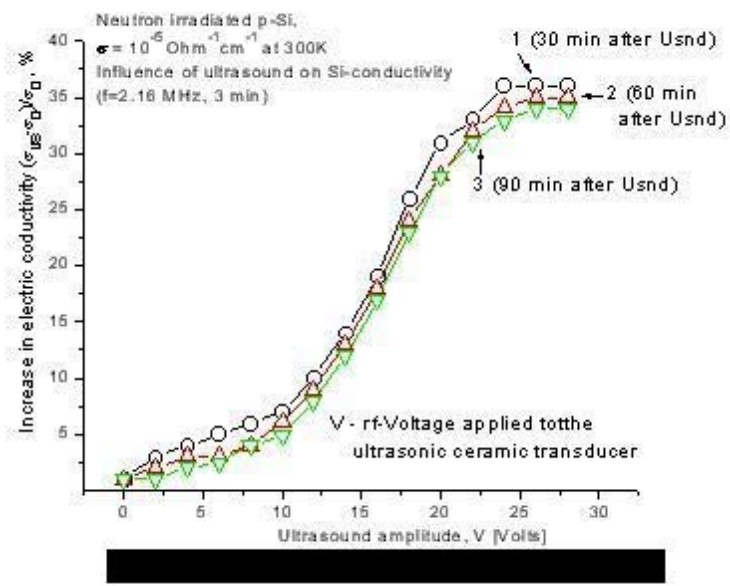

Fig. 1. Influence of UDM on electrical conductivity taken from a neutron irradiated
CZ-p-type silicon. Plot 1 – measurements are made 30 min after UDM, 2 – 60 min
after UDM, 3 – 90 min after UDM.

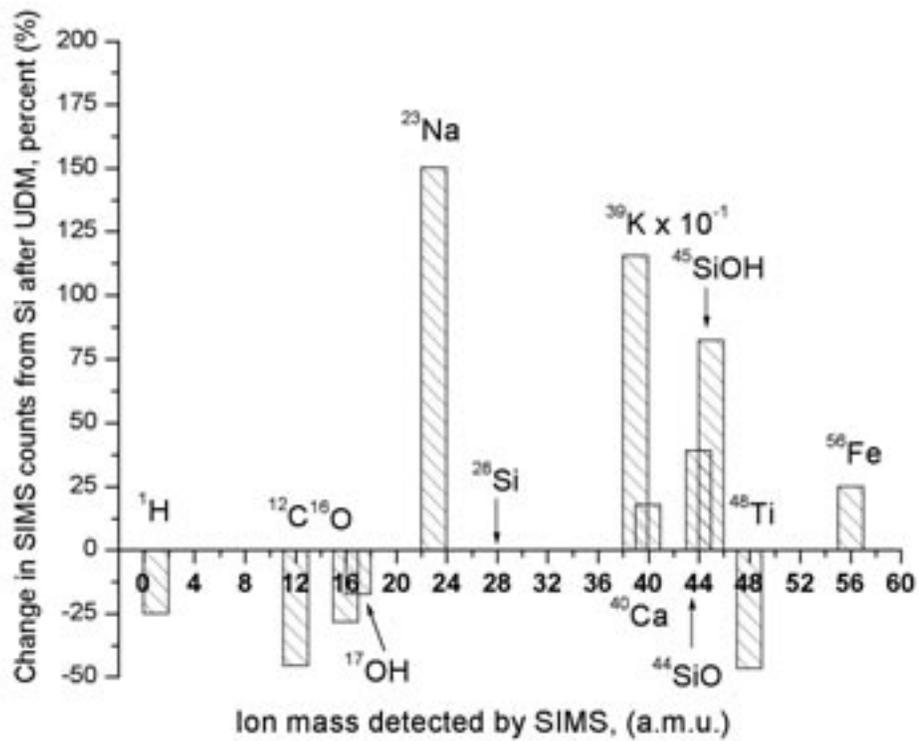

Fig. 2. Effect of UDM on SIMS spectrum from Cz-Si (sample Cz-p-Si-2K).
Negative values mean that the counts became lower after UDM, positive
values – counts became higher after UDM.